\documentclass[reprint, aps,prb,superscriptaddress, twocolumn, footinbib, longbibliography]{revtex4-1}

\usepackage{amsmath}
\usepackage{graphicx}
\usepackage{dcolumn}
\newcolumntype{d}[1]{D{.}{.}{#1}}

\usepackage{bm}
\usepackage{epstopdf}
\usepackage{graphicx}
\usepackage{stmaryrd}
\usepackage{tikz}
\usepackage{lipsum}
\usepackage{pgf}
\usepackage{float}
\usepackage[colorlinks, linkcolor= blue, citecolor = blue, urlcolor=blue]{hyperref}
\usepackage{tabularx}
\usepackage[none]{hyphenat}
\usepackage{subfigure}

\definecolor{lightblue}{RGB}{0,170,255}

\newcommand{\appropto}{\mathrel{\vcenter{
			\offinterlineskip\halign{\hfil$##$\cr
				\propto\cr\noalign{\kern2pt}\sim\cr\noalign{\kern-2pt}}}}}

\usepackage[normalem]{ulem}
\newcommand{\dps}{\displaystyle}

\newcommand{\om}{\iffalse}
\newcommand{\pd}[2]{\frac{\partial #1}{\partial #2}}

\newcommand{\ba}{\arraycolsep 0.3ex \begin{array}{rl}}
	\newcommand{\ea}{\end{array}}
\newcommand{\bc}{\begin{cases}}
	\newcommand{\ec}{\end{cases}}
\usepackage{color}
\newcolumntype{C}[1]{>{\centering\arraybackslash}p{#1}}

\begin{document}
	\title{Quantum Kinetic Theory of Nonlinear Optical Currents: Finite Fermi surface and Fermi sea contributions}
\author{Pankaj Bhalla}
    \email{pankaj.b@srmap.edu.in}
	\affiliation{Department of Physics, School of Engineering and Sciences, SRM University AP, Amaravati, 522240, India}
	\affiliation{ARC Centre of Excellence in Future Low-Energy Electronics Technologies, Australia}
\author{Kamal Das}
	\email{kamaldas@iitk.ac.in}
	\affiliation{Department of Physics, Indian Institute of Technology Kanpur, Kanpur-208016, India}
\author{Amit Agarwal}
	\email{amitag@iitk.ac.in}
	\affiliation{Department of Physics, Indian Institute of Technology Kanpur, Kanpur-208016, India}
\author{Dimitrie Culcer}
\email{d.culcer@unsw.edu.au}
	\affiliation{School of Physics, The University of New South Wales, Sydney 2052, Australia}
 	\affiliation{ARC Centre of Excellence in Future Low-Energy Electronics Technologies, The University of New South Wales, Sydney 2052, Australia}
\date{\today}

\begin{abstract}
The quantum kinetic framework provides a versatile method for investigating the dynamical optical and transport currents of crystalline solids. In this paper, starting from the density-matrix equations of motion, we present a general theoretical path to obtain the nonlinear optical response in an elegant and transparent manner. We devise an extensive kinetic theory that can be applied to materials with arbitrary band structures and captures intraband and interband coherence effects, finite Fermi surfaces, and disorder effects. We present a classification of the nonlinear optical currents arising from the interference of the interband and intraband components of the density matrix with distinct symmetry and quantum geometrical origin for each contribution. In this context, we report the following four primary findings. (i) The Fermi Golden Rule approach is insufficient to derive the correct expression for the injection current, a shortcoming that we remedy in our theory while associating the injection current with the intraband-interband contribution to the second-order density matrix. (ii) The interband-intraband contribution yields a resonant current that survives irrespective of any symmetry constraint in addition to the well-known anomalous nonlinear current (non-resonant), which requires time-reversal symmetry. (iii) Quite generally, the nonlinear current is significantly enhanced by contributions arising from the finite Fermi surface. (iv) The finite Fermi surface and Fermi sea additionally lead to sizable novel nonlinear effects via contributions we term double resonant and higher-order pole. We investigate such effects in sum frequency and difference frequency generation. As an illustration, we compute the nonlinear response of the topological antiferromagnet CuMnAs and thin film tilted Weyl semimetals as model systems dominated by interband coherence contributions. We find that the nonlinear response of CuMnAs is responsive to the direction of the finite magnetization field and the response of Weyl semimetal to the tilt. In addition, the choice of the polarization angle of the beam is crucial to have a nonlinear current in CuMnAs, while it is not the case for Weyl semimetals. 
\end{abstract}

\maketitle
\section{Introduction}
Probing novel nonlinear optical effects due to the light-matter interaction has recently become a subject of great interest due to the cutting-edge advancement in fields such as ultra-fast phenomenon and optoelectronics \cite{boyd_book, rubin_PR1962, morimoto_SA2016, Orenstein_ARCM2021}. In particular, nonlinear optical responses are essential to understanding the symmetry and geometry of the electron wave function. Nonlinear phenomena attracting recent interest include second and third harmonic generation, rectification, and photocurrents, all of which are intimately tied to the nature of the Bloch wave functions \cite{fiebig_JOSAB2005, dean_APL2009, dean_PRB2010, mikhailov_PRB2011, kaminski_PRL2009, mciver_PRB2012, you_NP2019, takasan_PRB2021, li_PRB2018, hipolito_2DM2017, Sipe_PRB1993, sipe_PRB2000_second, zhang_PRL2019, gao_PRB2021, fei_PRB2020, nagaosa_NC2018, bhalla_PRB2022}. Specifically, these effects are mainly determined by the momentum-space quantum geometric quantities such as Berry curvature, quantum metric, and metric connection, which tend to be large in systems having broken inversion symmetry, time-reversal symmetry, or both \cite{xiao_RMP2010, Ahn_NP2022, min_NSR2019, ahn_PRX2020}. 

One famous example that has been studied for decades is the anomalous Hall effect due to the finite Berry curvature in time-reversal symmetry broken systems \cite{nagaosa_RMP2010, culcer_2022, culcer_PRB2003}. Such responses are robust in gapped systems, which reflect the topology of the system \cite{yu_Sc2010, chang_Sc2013, cullen_PRL2021}. In addition to these, recently, the local quantum geometrical quantity -- the Berry curvature -- has been shown to be a key driver of second-order responses such as the linear photo-currents, quantized circular photo-currents, injection currents, and the nonlinear Hall effect \cite{sipe_PRB2000_second, hosur_PRB2011, tan_PRB2019, sodemann_PRL2015, morimoto_PRB2016, rostami_PRB2018_nonlinear, cong_PRB2019, ahn_PRX2020_low, watanabe_PRX2021, bhalla_PRL2021a, kaplan_arxiv2022, glazov_PRL2020, gao_PRB2021, zeng_PRB2021, sinha_NP2022_berry, Chakra_2Dmat2022_nonlinear, kumar_PRB2020,lahiri_PRB2021_nonlinear,Lahiri_INHE_2022,Harsh_2022, bhalla_PRB2021, zhang_PRB2022, kaplan_PRR2022, golub_PRB2022, leppenen_PRB2022}. An example of a linear photo-current is the shift current, which reflects the shift in the position of the electron wave packet upon excitation from the valence to the conduction band, and is observed in bismuth telluride, where it is related to Fermi surface anisotropy induced by warping effects \cite{kim_PRB2017}. In contrast, in centrosymmetric systems, the shift current is contributed by photon drag processes arising from non-vertical transitions \cite{shi_PRL2021}.
Second harmonic generation (SHG) is another example of a second-order response with a geometrical origin: experimentally it has been observed in inversion symmetry breaking TaAs Weyl semimetals, whose Weyl nodes are monopoles of the Berry curvature \cite{wu_NP2017, patankar_PRB2018, sirica_PRL2019}. All these second-order photocurrents and SHG phenomena come under the roof of most general second-order frequency conversion phenomena such as sum frequency generation (SFG), and difference frequency generation (DFG) \cite{shen_book, morita_book, juan_PRX2020}. Conventionally, these processes arise due to the interference of two incoming beams having frequencies $\omega_i$ and $\omega_j$, which generate one outcoming radiation with net frequency $\omega_{\delta} = \omega_i \pm \omega_j$. The signal associated with the SFG is governed by third rank tensor $\sigma_{abc}(\omega_{\delta}; \omega_j, \omega_l)$, and the DFG by $\sigma_{abc}(\omega_{\delta}; \omega_j, -\omega_l)$. We explore both of these in the present study. 

Remarkably, most studies have been concerned with nonlinear processes specific to particular materials and in a narrow region of applicability -- overwhelmingly focussing on clean, undoped materials \cite{sipe_PRB2000_second, Sipe_PRB1993, Aversa_PRB1995}. Nevertheless, in doped systems, the limit $1/\tau \rightarrow 0$, with $\tau$ the relaxation time scale, is unrealistic. Although nonlinear currents stemming from the Fermi sea, such as shift, and injection, give finite results, recently explored nonlinear currents induced by finite Fermi surface contributions, such as Drude, resonant photogalvanic, and double resonant currents diverge \cite{bhalla_PRL2020, bhalla_PRL2022}. Thus to understand the overall behavior of quantum geometry-driven currents in a consistent manner, it is important to develop a formalism that comprehensively accounts for nonlinear optical currents with distinct physical origins within both clean and dirty limits, which in our opinion, is lacking in the literature.

\begin{figure*}[t]
    \centering
    \includegraphics[width=14cm]{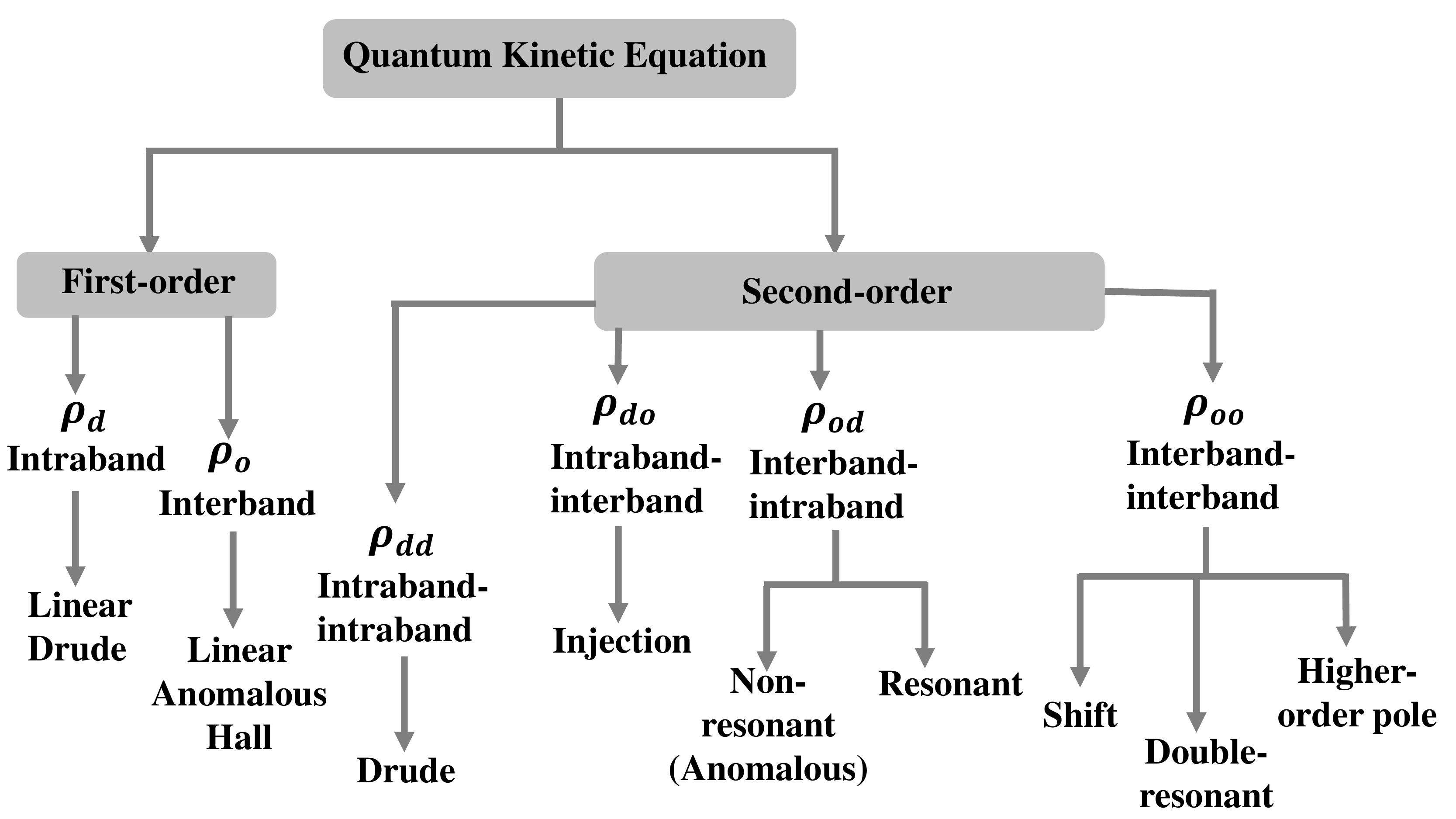}
    \caption{A schematic tree for the generation of different contributions of the first-order and second-order density matrix $\rho$ that leads to distinct forms of nonlinear currents. Here the subscripts $d$ and $o$ stand for the diagonal and off-diagonal parts of the density matrix. In the double subscripts such as $dd$, $do$, $od$ and $oo$, the first letter indicates the diagonal and off-diagonal part of the second-order density matrix and later letter corresponds to the dependence of the relevant part of the first-order density matrix on the second-order.}
    \label{fig:schematic}
\end{figure*}

In this paper, we systematically elaborate the dynamics of the nonlinear optical current in response to the light-matter interaction in length gauge within the density matrix formalism while accounting for the disorder and Fermi surface and sea effects in doped systems. We provide the general framework of the quantum kinetic theory for distinct second-order optical processes in response to the external electric or laser fields. We focus on the second-order currents arising from the mutual interference of the intraband and interband effects. This allows us to classify the nonlinear current, according to its origins, into four types, namely intraband-intraband, intraband-interband, interband-intraband, and interband-interband, as shown schematically in the tree map of currents in Fig.~\ref{fig:schematic}. The Drude current or the intraband-intraband current is generated due to the momentum derivative of the non-equilibrium distribution function and shows a divergence in the clean limit. The intraband-interband arises through the velocity difference between the distinct bands and is known as the injection current. The contribution of the intraband response to the interband part of the density matrix leads to the nonlinear anomalous current proportional to the Berry curvature, which survives only for a time-reversal symmetric system and shows a non-resonant structure. However, the resonant counterpart is proportional to the quantum metric and contributes to $\mathcal{P}\mathcal{T}$ symmetric systems, which will constitute an essential segment of this paper. The complete interband coherence effect yields the shift current due to the shift of the wave packet, double resonant due to the asymmetric Fermi surface and higher-order pole by the momentum displaced joint density of states. In addition, we observe that the finite Fermi surface generates strong absorption peaks in the resonant (part of interband-intraband) and double resonant (part of interband-interband) components, thus enhancing the total nonlinear current. 

Within the quantum kinetic approach, our study reveals that the finite Fermi surface contribution is key to the \textit{resonant} nature of nonlinear responses in doped systems. This originates from distinct quantum geometric quantities. In CuMnAs, for example, the nonlinear response is sensitive to the direction of the magnetization field. Likewise, the nonlinear current due to linearly polarized light depends on the propagation direction of the incident beams. The second-order current with the magnetization direction along $\hat{x}$-axis varies as $j_y^{(2)} \sim \cos^2\gamma$ and $j_x^{(2)} \sim \sin2\gamma$, where $\gamma$ is the polarization angle along $\hat{x}$-axis. On the other hand, in thin films of tilted Weyl semimetals, the quantum geometric quantities are insensitive to the tilt, which is the parameter breaking time-reversal. At the same time, the finite tilt makes the response more pronounced. In addition, the nonlinear current is generated here, irrespective of the choice of the polarization angles. Moreover, we discuss in detail how the different nonlinear response components contribute to optical currents using the symmetry properties of quantum geometric quantities and measurement geometry.


The paper is organized as follows. In Sec.~\ref{sec:optical}, the general theoretical kinetic framework to compute the optical currents is presented. Here, we solve the kinetic equations in momentum space to calculate the components of the density matrix, both diagonal and off-diagonal, in the band index for the linear and nonlinear cases. Next, the optical nonlinear currents are calculated by employing the diagonal and off-diagonal density matrix solutions. In Sec.~\ref{sec:symmetry}, the symmetry analysis of different quantum geometric quantities is given, and the spatial geometrical analysis for various optical conductivity tensor components is discussed. Later, the theory is tested for materials such as topological antiferromagnetic CuMnAs and thin film tilted Weyl semimetal in Sec.~\ref{sec:applications}. Finally, in Sec.~\ref{sec:summary}, we conclude with future perspectives. 

\section{Optical Currents and Quantum Kinetic Approach}
\label{sec:optical}

In this section, we determine the general form of the nonlinear currents in response to the optical electric field. Specifically, we derive the polarization or dipole moment per unit volume of a system depending on the strength of the optical field. Phenomenologically, in response to an optical field, the time-dependent polarization for a lossless medium can be expressed in the form \cite{boyd_book, bloembergen_book}
\begin{equation}
\tilde P_{a}(t) = \sum_{b} \chi_{ab} \tilde E_{b}(t) + \sum_{bc} \chi_{abc} \tilde E_{b}(t)\tilde E_{c}(t) + \cdots,
\end{equation}
where $\chi_{ab}$, and $\chi_{abc}$ are optical susceptibilities of second-rank and third-rank respectively and $\tilde{\bm E}(t)$ is the optical field. Further, using the relation between the polarization and the current ${\bm j} = d{\bm P}/dt$, the optical current is written in the form
\begin{equation}
    \tilde{\bm j}(t) = \tilde{\bm j}^{(1)}(t) + \tilde{\bm j}^{(2)}(t) + \cdots 
\end{equation}
Here, $\tilde{\bm j}^{(1)}$ and $\tilde{\bm j}^{(2)}$ are proportional to the first and second power of the optical field, respectively. For an optical field of the form $\tilde{\bm E}(t)= \sum_j {\bm E}^{\omega_j} e^{-i \omega_j t}$ where ${\bm E}^{-\omega_j}={\bm E}^{*\omega_j}$ having ${\bm E}^*$ as the complex conjugate of the field ${\bm E}$, the second-order optical current can be written as 
\begin{equation}
    \tilde j_a^{(2)} (t) = \sum_{bc}  \sigma_{abc} \tilde E_b(t) \tilde E_c(t),
\end{equation}
where $\sigma_{abc}$ denotes the second-order optical conductivity of the system.
Further, depending on the frequency components of the optical field, the nonlinear optical current is given by
\begin{equation}
\tilde j_a^{(2)} (t) = \sum_{\delta} j_a^{(2)}(\omega_\delta) e^{-i\omega_\delta t},
\end{equation}
where the summation runs over distinct components of the frequency. The different combinations of frequency components lead to various nonlinear phenomena, namely sum frequency summation (SFG) $j_a^{(2)}(\omega_j+\omega_l)$, second harmonic generation (SHG) $j_a^{(2)}(2\omega_j)$, difference frequency generation (DFG) $j_a^{(2)}(\omega_j - \omega_l)$, and optical rectification (OR) $j_a^{(2)}(0)$. We employ the density matrix approach to evaluate the corresponding nonlinear response functions, which is discussed in the following subsection.

\subsection{Density matrix approach}
We start from the quantum Liouville equation for the time-dependent single-particle density matrix $\rho ({\bm k},t)$ in the momentum space \cite{culcer_PRB2017},
\begin{equation} \label{eqn:DM1}
\ba
&\dps \pd{\rho ({\bm k},t)}{t} + \frac{i}{\hbar}[\mathcal{H}({\bm k},t), \rho ({\bm k},t)] =0.
\ea
\end{equation}
Here $\mathcal{H}({\bm k},t)$ is the full Hamiltonian of the system, including the light-matter interaction, and $[\cdot, \cdot]$ refers to the commutator bracket. In the length gauge, the perturbed Hamiltonian for the spatially uniform and time-varying optical field reduces in the form
\begin{equation} \label{eqn:Ham}
\mathcal{H}({\bm k},t) = \mathcal{H}_0({\bm k}) + \mathcal{H}_E(t) + U,
\end{equation}
where $\mathcal{H}_0({\bm k})$ is the unperturbed and Bloch Hamiltonian of the system, $\mathcal{H}_E(t) = e {\bm r} \cdot {\bm E}(t)$ represents the interaction with the electric field, `$-e$' the electronic charge and $U$ the disorder potential. Considering Eq.~\eqref{eqn:Ham}, the quantum kinetic equation \eqref{eqn:DM1} on averaging over disorder configurations takes the form
\begin{equation}
\pd{\rho({\bm k},t)}{t} + \frac{i}{\hbar} [\mathcal{H}_0,\rho({\bm k},t)] +  J(\rho({\bm k},t)) = -\frac{i}{\hbar} [\mathcal{H}_E,\rho({\bm k},t)].
\label{eqn:dens}
\end{equation}
Here $J(\rho)$ represents the scattering term that takes into account the impact of the disorder potential. In the present framework, we treat the scattering term under the relaxation time approximation and approximate the term as $\rho({\bm k},t)/\tau$, with $\tau$ being a parameter specifying the time taken to relax the system towards the equilibrium state or the relaxation time scale. For simplicity, we consider the relaxation time scale $\tau$ as a constant parameter across the Fermi surface. Therefore, the kinetic equation becomes
\begin{equation}
\pd{\rho}{t} + \frac{i}{\hbar} [\mathcal{H}_0,\rho] +  \frac{\rho - \rho^{(0)}}{\tau} = -\frac{i}{\hbar} [\mathcal{H}_E,\rho].
\label{eqn:den}
\end{equation}
We have written $\rho({\bm k},t)$ as $\rho$ to simplify the notation, while $\rho^{(0)}$ is the equilibrium density matrix. To find the solution for the kinetic equation, we expand the density matrix perturbatively in the powers of the time-dependent and space-homogeneous optical field.
\begin{equation}
\rho = \rho^{(0)} + \rho^{(1)} + \rho^{(2)} + \cdots,
\end{equation}
where $\rho^{(N)} \propto E^{(N)}$ having the superscript `$N$' for an order of the field. 

In the band basis representation, the density matrix for the most simple two-band model can be represented as
\begin{equation}
    \rho = \begin{pmatrix}
    \rho_{mm} & \rho_{mp}\\
    \rho_{pm} & \rho_{pp}
    \end{pmatrix},
\end{equation}
where $m$ and $p$ refer to band indices. Within this band basis representation, Eq.~\eqref{eqn:den} for the $N^{\text{th}}$ order density matrix can be written
\begin{equation}
\pd{\rho_{mp}^{(N)}}{t} + \frac{i}{\hbar} [\mathcal{H}_0,\rho^{(N)}]_{mp} +  \frac{\rho^{(N)}_{mp}}{\tau} = \frac{e{\bm E}(t)}{\hbar} \cdot \left[ D_{\bm k} \rho^{(N-1)} \right]_{mp}.
\label{eqn:den1}
\end{equation}
The covariant derivative~\cite{nagaosa_AM2017_concept} $[D_{\bm k} \rho]_{mp} = \partial_{\bm k} \rho_{mp} -i[\mathcal{R}_{\bm k}, \rho]_{mp}$ where $\mathcal{R}_{mp}({\bm k}) = \langle u_{\bm k}^{m} \vert i \partial_{\bm k}u_{{\bm k}}^{p} \rangle$ is the momentum space Berry connection with $\vert u_{\bm k}^{m} \rangle$ the periodic part of the Bloch wave-function, and $\partial_{\bm k}$ represents the momentum derivative. This is obtained by inserting the expression for $\mathcal{H}_E$ in the commutator $[\mathcal{H}_E,\rho]$ and using the relation $| m,{\bm k} \rangle = e^{-i{\bm k}\cdot {\bf r}} |u_{\bm k}^{m}\rangle$ and $\hat{{\bm r}}| m,{\bm k} \rangle = i [\partial_{\bm k}e^{-i{\bm k}\cdot {\bf r}}] |u_{\bm k}^{m}\rangle $. Further, the corresponding right-hand side term serves as the driving term that generates a response in the system. It is also a fully intrinsic term and is determined by the electronic structure of the system. Note that the right-hand side of the equation contains the $(N-1)^{th}$-order density matrix due to the presence of the field factor. It is also evident from this expression that to find the solution of the density matrix of order $N \ge 1$, one requires the solution for the proceeding order due to $(N-1)$ order term in the right-hand side of the Eq.~\eqref{eqn:den1}. However, for $N=0$ case the right side of Eq.~\eqref{eqn:den1} approaches to zero which gives $\rho_{mp}^{(0)} = f^0(\varepsilon_{\bm k}^m) \delta_{mp}$ the equilibrium Fermi-Dirac distribution function and is defined as $f^0(\varepsilon_{\bm k}^m)  = [e^{\beta(\varepsilon_{\bm k}^m-\mu)} +1]^{-1}$ having $\beta = [k_B T]^{-1}$ with $k_B$ the Boltzmann constant, $T$ the electron temperature, $\mu$ represents the chemical potential, and $\varepsilon_{\bm k}^m$ corresponds to the electron dispersion for $m^{\text{th}}$ band. To study the dynamics of the linear and nonlinear currents, we calculate the diagonal ($m=p$) or the intraband and the off-diagonal ($m \neq p$) or the interband part of the density matrix correspond to the linear and the quadratic power of the electric field in the following subsections.
\subsubsection{Linear order density matrix}
To recover the linear response, we set $N=1$ and solve the kinetic equation by splitting the density matrix into diagonal and off-diagonal components in the band index such as $\rho = \rho_{mm}\delta_{mp} +  \rho_{mp}$. Here the first term is the diagonal or intraband part of the density matrix owing to the Dirac delta function, which vanishes for insulators. The second term refers to the off-diagonal or interband
coherence part of the density matrix. 

Firstly for the intraband contribution to the density matrix or $m=p$ case, Eq.~\eqref{eqn:den1} reduces to
\begin{equation}
\pd{\rho_{mm}^{(1)}}{t} +  \frac{\rho_{mm}^{(1)}}{\tau_{\alpha}} = \frac{e{\bm E}(t)}{\hbar} \cdot \pd{ f^0(\varepsilon_{\bm k}^m) }{{\bm k}},
\label{eqn:den2}
\end{equation}
where $\tau_{\alpha}$ is the time scale for the intraband transitions. It is to be noted that the commutator between the Bloch state Hamiltonian $\mathcal{H}_0$ and the zeroth order density matrix $\rho^{(0)}$ is zero. On solving the linear order differential equation by taking an integrating factor, the intraband time-dependent density matrix takes the form 
\begin{equation}
\rho_{mm}^{(1)} =\frac{e}{\hbar} \sum_{j} \partial_c f_{m}^{0}  g_{0;\alpha}^{\omega_j} E_c^{\omega_j} e^{-i\omega_j t}. 
\label{eqn:den3}
\end{equation}
Here, we define $g_{0;\alpha}^{\omega_j} = [1/\tau_\alpha - i\omega_j]^{-1}$ and $\partial_{k_c} \equiv \partial_{c}$ for brevity. Similarly, the interband component ($m \neq p$) of the density matrix comes out to be
\begin{equation} \label{eqn:den4}
\pd{ \rho_{mp}^{(1)}}{t} + i\omega_{mp}\rho_{mp}^{(1)} + \dfrac{\rho_{mp}^{(1)}}{\tau_\gamma}=i \frac{e E_c(t)}{\hbar}  {\mathcal R}_{mp}^c F_{mp}.
\end{equation}
With $\hbar \omega_{mp} = \varepsilon_{m,{\bm k}}-\varepsilon_{p,{\bm k}}$ as the interband transition energy at momentum ${\bf k}$, $\tau_\gamma$ as the time scale corresponding to the interband transitions and $F_{mp}=f^0(\varepsilon_{\bm k}^m) - f^0(\varepsilon_{\bm k}^p)$ is the difference in the occupation between two distinct bands.
Here we use the relation $D_{\bm k}\rho^{(0)} = -i[\mathcal{R}_{\bm k}, \rho^{(0)}]$ due to vanishing $\partial_{\bm k}\rho_{mp}^{(0)}$ as the equilibrium part contains only the diagonal elements in the band index.
The solution of Eq.~\eqref{eqn:den4} for the off-diagonal part of the density matrix gives
\begin{equation}
\rho_{mp}^{(1)} = i\frac{e}{\hbar} \sum_{j} \mathcal{R}_{mp}^{c} F_{mp}  g_{mp;\gamma}^{\omega_j} E_c^{\omega_j} e^{-i\omega_j t},
\label{eqn:den4a}
\end{equation}
where $g_{mp;\gamma}^{\omega_j} = [1/\tau_\gamma - i(\omega_j-\omega_{mp})]^{-1}$ relates to the joint density of states broadened by the interband relaxation time scale.
From this, we find that the linear order $\rho_{mp}^{(1)}$ depends on the shift of the Fermi function $F_{mp}$ and this contribution survives only for $m \neq p$. In a compact form, the complete solution of the first-order density matrix can be written as
\begin{equation}
\rho_{mp}^{(1)} = \dfrac{e}{\hbar} \sum_j   \tilde \rho_{mp;j}^{(1), c} E_c^{\omega_j}  e^{-i\omega_j t},
\label{rho_mp_1_f}
\end{equation}
where 
\begin{equation}
\ba
\tilde \rho_{mp;j}^{(1), c} = \tilde \rho_{mp}^{(1), c}(\omega_j) &\dps =  \partial_{c} \rho_{mp}^{(0)} g_{0;\alpha}^{\omega_j} \delta_{mp}+ i{\mathcal R}_{mp}^c F_{mp}g_{mp;\gamma}^{\omega_j}. 
\ea
\end{equation}
This expression can be further simplified in the low temperature region by replacing the Fermi distribution function and its energy derivative with the Heaviside step function $\Theta(\varepsilon_{\bf k}^m - \mu)$ and Dirac delta function $-\delta(\varepsilon_{\bf k}^m - \mu)$.

\subsubsection{Density matrix to second-order in the electric field}

Substituting $N=2$ in Eq.~\eqref{eqn:den1}, we get
\begin{equation}
\pd{ \rho_{mp}^{(2)}}{t} + \dfrac{i}{\hbar}[\mathcal{H}_0 ,\rho^{(2)} ]_{mp} + \dfrac{\rho_{mp}^{(2)}}{\tau}= \frac{e E_b(t)}{\hbar} [D_b \rho^{(1)}]_{mp}.
\end{equation}
Using the solution for the linear order density matrix Eq.~\eqref{rho_mp_1_f}, we obtain
\begin{equation}
\ba
&\dps \pd{ \rho_{mp}^{(2)}}{t} + i\omega_{mp}\rho^{(2)}_{mp} + \dfrac{\rho_{mp}^{(2)}}{\tau_\alpha} 
\\[3ex]
&\dps = \frac{e^2 }{\hbar^2} \sum_{ P}\sum_{j,l}\bigg\{ \left[ \partial_b \tilde \rho_{mp;l}^{(1), c}  -i \sum_{n} \left( \mathcal{R}^{b}_{mn} \tilde \rho_{np;l}^{(1), c} - \tilde \rho_{mn;l}^{(1), c} \mathcal{R}^{b}_{np} \right)\right] 
\\[3ex]
&\dps ~~~~ E_b^{\omega_l} E_c^{\omega_j}   e^{- i(\omega_j + \omega_l) t} \bigg\}.
\ea
\end{equation}
Here, the sum over $P$ refers to the intrinsic permutation symmetry $(b , \omega_l \leftrightarrow c, \omega_j)$.
For $m=p$ case, $\rho_{mm}^{(2)}$ can be written as a sum of two terms,
$\rho_{mm}^{(2)} = \rho_{mm}^{(2),dd} + \rho_{mm}^{(2),do}$ where the first term refers to the intraband-intraband (or diagonal-diagonal) part and stems from the diagonal component of the first-order density matrix. The second term corresponds to the intraband-interband (or diagonal-off diagonal) part by the off-diagonal component of the first-order density matrix. The first term $\rho_{mm}^{(2),dd}$ is defined as
\begin{equation} \nonumber
\label{eqn:dd}
\rho_{mm}^{(2),dd} = \dfrac{e^2}{\hbar^2} \sum_{\mathcal{P}}\sum_{j,l}   \tilde \rho_{mm}^{(2),dd}(\omega_{\delta};\omega_j,\omega_l) E_b^{\omega_l} E_c^{\omega_j}  e^{-i\omega_\delta t},
\end{equation}
having
\begin{equation}
\tilde \rho_{mm}^{(2),dd} (\omega_\delta;\omega_j,\omega_l) = \partial_b \tilde \rho_{mm;l}^{(1), c}    g_{0;\alpha}^{\omega_\delta}.
\end{equation}
Here, we define $\omega_{\delta} = \omega_j + \omega_l$. Similarly, the intraband-interband part of the density matrix $\rho_{mm}^{(2),do}$ is given by
\begin{equation}\label{eqn:od}
\ba
\tilde \rho_{mm}^{(2),do}(\omega_{\delta};\omega_j,\omega_l)  =&\dps  -i \sum_{n} \left( \mathcal{R}^{b}_{mn} \tilde \rho_{nm;l}^{(1), c} - \tilde \rho_{mn;l}^{(1), c} \mathcal{R}^{b}_{nm} \right) g_{0;\alpha}^{\omega_{\delta}}.
\ea
\end{equation}
In the same spirit, the density matrix for $m \neq p$ case can be expressed in the form $\rho_{mp}^{(2)} = \rho_{mp}^{(2),od} + \rho_{mp}^{(2),oo}$.
Firstly, the interband-intraband part $\rho_{mp}^{(2),od}$ is defined in the form
\begin{equation}
\label{eqn:do}
\tilde \rho_{mp}^{(2)od}(\omega_{\delta};\omega_j,\omega_l) =  -i \mathcal{R}^{b}_{mp} \left(  \tilde \rho_{pp;l}^{(1), c} - \tilde \rho_{mm;l}^{(1), c}  \right) g_{mp;\gamma}^{\omega_\delta}.
\end{equation}
Secondly, the interband-interband part of the density matrix is given by 
\begin{equation}\label{eqn:oo}
\ba
 \tilde \rho_{mp}^{(2),oo} (\omega_{\delta};\omega_j,\omega_l) =  &\dps g_{mp;\gamma}^{\omega_\delta}\Big[ \left\{ \partial_b  -i  \left( \mathcal{R}^{b}_{mm}  - \mathcal{R}^{b}_{pp} \right) \right\} \tilde \rho_{mp;l}^{(1), c} 
\\[3ex]
&\dps -i \sum_{n \neq (m,p)} \left( \mathcal{R}^{b}_{mn} \tilde \rho_{np;l}^{(1), c} - \tilde \rho_{mn;l}^{(1), c} \mathcal{R}^{b}_{np} \right) \Big].
\ea
\end{equation}
Here, the second term in the open brackets contributes to multi-band systems. 

\subsection{Nonlinear Optical Currents}
To obtain the expressions for nonlinear (i.e., second-order in the field) optical currents, we use the general definition of the current, which is the trace of the velocity operator with the density matrix. Accordingly, the time-dependent optical current for $d$-dimensional momentum space in the band index basis can be expressed in the form
\begin{equation}
{\bm j}(t)  = -e \sum_{p} \int \frac{d^d{\bm k}}{(2\pi)^d} \langle p \vert {\bm v}({\bm k}) \rho({\bm k},t)\vert p \rangle.
\end{equation}
Here, ${\bm v}_{pm}({\bm k})$ the velocity matrix in the band basis in the Bloch representation is ${v}^a_{pm}({\bm k}) = \hbar^{-1} (\delta_{pm} \partial_{a}\varepsilon_{\bm k}^{m} + i  \mathcal{R}^a_{pm} \hbar\omega_{pm})$. The first part refers to the group velocity of Bloch electrons along the spatial direction $a$, and the second part corresponds to the interband velocity component.
Using this, the current in an arbitrary direction can be expressed as $j^a(t)= -e \sum_{m,p} \sum_{\bm k}  v_{pm}^{a} \rho_{mp}^{(2)}(t)$. Below we make use of the different interband and intraband components of the density matrix, such as $\rho^{dd}$, $\rho^{do}$, $\rho^{od}$ and $\rho^{oo}$ to obtain the nonlinear currents and classified them accordingly.
\subsubsection{Intraband-Intraband Current}
For the intraband-intraband ($dd$) contribution of the density matrix, the current is given by 
\begin{equation}
    \tilde j_a^{(2),dd}(t) = -e \sum_{m}\sum_{\bm k} v_{mm}^{0a} \rho_{mm}^{(2),dd}(t),
\end{equation}
where $v_{mm}^{0a} = \hbar^{-1} \partial_{a}\varepsilon_{\bm k}^{m}$ is the diagonal component of the velocity in the band basis. Using Eq.~\eqref{eqn:dd} in the above equation, the second-order conductivity $\sigma_{dd}$ can be expressed in the form
\begin{equation}
\sigma^{dd}_{abc}(\omega_{\delta}; \omega_j, \omega_l) = - \frac{e^3 }{\hbar^2} \sum_{\mathcal{P}}\sum_{m} v_{mm}^{0a} \partial_b \partial_c f_{m}^{(0)} g_{0;\alpha}^{\omega_l} g_{0;\alpha}^{ \omega_{\delta}}.
\label{j_Drude}
\end{equation}
It is clear from the expression that $\sigma^{dd}$ is the Fermi surface contribution due to the presence of the momentum derivative of the Fermi function. In addition, it is entirely the intraband contribution and stems from the single band only. Note that this contribution to the current is only finite for metals and semimetals and can be completely ignored for insulators.

%
Moreover, the frequency dependence of this optical current is captured by the product of two $g_0$ factors having the following form
\begin{equation}
\label{eqn:gD}
g_{0;\alpha}^{\omega_l} g_{0;\alpha}^{\omega_{\delta}} = \dfrac{1}{1/\tau_\alpha -i \omega_l }\times  \dfrac{1}{1/\tau_\alpha -i  \omega_{\delta}}.
\end{equation}
It is helpful to expand the product of $g_0$ factors by partial separation below as
\begin{equation}
g_{0;\alpha}^{\omega_l} g_{0;\alpha}^{ \omega_{\delta}} = \frac{\tau_\alpha}{i(\omega_l -  \omega_{\delta})} \left(\frac{1-i\omega_l\tau_\alpha}{1+\omega_l^2 \tau_\alpha^2} - \frac{1-i \omega_{\delta} \tau_\alpha}{1+ \omega_{\delta}^2\tau_\alpha^2} \right),
\end{equation}
where we have considered $(\omega_l - \omega_\delta) \neq 0$.
We emphasize that one can not reproduce the dc limit from the above equation.
In the low frequency limit, quantitatively defined as $\omega \tau \ll 1$, Eq.~\eqref{eqn:gD} reduces as $   g_{0;\alpha}^{\omega_l} g_{0;\alpha}^{\omega_{\delta}} = \tau_\alpha^2$,
which has the Drude nonlinear current like relaxation time dependence. In the opposite frequency regime $\omega\tau \gg 1$, Eq.~\eqref{eqn:gD}, the product of $g$-factors varies with the inverse quadratic dependence of the frequency like $g_{0;\alpha}^{\omega_l} g_{0;\alpha}^{\omega_{\delta}} = 1/\omega_{\delta}\omega_l$. This generates a similar expression as quoted in Ref.~[\onlinecite{watanabe_PRX2021}] and is referred to as the nonlinear Drude term. Note that the nonlinear Drude current does not rely on band geometric quantities.
\subsubsection{Intraband-Interband Current}
We consider the contribution of the intraband-interband ($do$) part of the density matrix $\rho_{mm}^{(2),do}$ to the current. Here, the response arises from the band geometrical quantity and the change in group velocity of the carrier on transiting from one band to another band. The resulting nonlinear current is termed as the injection current. It is given by 
\begin{equation}
j_a^{(2),do}(t) = -e \sum_{m}\sum_{\bm k} v_{mm}^{0a} \rho_{mm}^{(2),do}(t).
\end{equation}
Using the nonlinear density matrix calculated in Eq.~\eqref{eqn:od}, we have the following expression for the nonlinear conductivity 
\begin{equation}
\sigma_{abc}^{do}(\omega_{\delta}; \omega_l,\omega_j) = \frac{e^3 }{\hbar^2} \sum_{\mathcal{P}}\sum_{m,p}  \Delta^a_{mp}  \mathcal{Q}_{mp}^{bc}    
F_{mp} g_{mp;\gamma}^{\omega_l}    g_{0;\alpha}^{\omega_{\delta}}.
\label{sigma_do}
\end{equation}
Here, we have defined $\Delta^a_{mp} = (v_{mm}^{0a}-v_{pp}^{0a})$, the difference between the group velocities of two bands. We have also defined the geometrical quantity, the band-resolved quantum geometric tensor (QGT)~\cite{ma_PRB2010_abelian} as ${\mathcal Q}_{mp}^{bc}=\mathcal{R}^{b}_{pm} \mathcal{R}^{c}_{mp} = (G_{mp}^{bc} - i/2 \Omega_{mp}^{bc})$.
Here, the quantum metric is $G_{mp}^{bc} = \{ {\mathcal R}^b_{pm}, {\mathcal R}_{mp}^c \}/2$ which is symmetric under the exchange of spatial and band indices. On the other hand, the Berry curvature $\Omega_{pm}^{bc} = i[{\mathcal R}^b_{pm}, {\mathcal R}_{mp}^c]/2$ is antisymmetric under the exchange of band and spatial indices. 
The optical frequency dependence of this part of conductivity comes from
\begin{equation}
g_{mp;\gamma}^{\omega_l} g_{0;\alpha}^{ \omega_\delta} = \dfrac{1}{1/\tau_\gamma -i( \omega_l - \omega_{mp}) }\times  \dfrac{1}{1/\tau_\alpha -i \omega_{\delta}}.
\end{equation}
On expanding the product of $g$-factors, one ends up with four terms having two real and two imaginary terms. Here, the leading contributing terms to the nonlinear current are
\begin{align} \nonumber
\label{eqn:GFs}
g_{mp;\gamma}^{\omega_l} g_{0;\alpha}^{ \omega_{\delta}} &\approx \frac{1/\tau_{\alpha}}{1/\tau_{\alpha}^2 + \omega_{\delta}^2} \frac{1/\tau_{\gamma}}{1/\tau_{\gamma}^2 + (\omega_l - \omega_{mp})^2} \\
&
+ 
i \frac{\omega_{\delta} }{1/\tau_{\alpha}^2 + \omega_{\delta}^2} \frac{1/\tau_{\gamma}}{1/\tau_{\gamma}^2 + (\omega_l - \omega_{mp})^2}.
\end{align}
This yields two terms to the interband-intraband conductivity. However, when we take the limit $\omega_{\delta} \rightarrow 0$ in the above expression and then consider $1/\tau_\gamma \rightarrow 0$, the second term vanishes. The resulting expression is
\begin{align} \nonumber
\label{eqn:Naga}
&\sigma_{abc}^{do} (\omega_{\delta}; \omega_l,\omega_j) = \\[1ex]  &-\frac{2\pi e^3 }{\hbar^2} \tau_\alpha \sum_{b,c} \sum_{m,p} \Delta_{mp}^{a} \mathcal{Q}_{mp}^{bc} F_{mp} \delta (\omega_{mp} - \omega_l).
\end{align}
Here we use $g_{mp;\gamma}^{\omega_l} g_{0;\alpha}^{ \omega_{\delta}} = - \tau_\alpha \pi \delta (\omega_{mp} - \omega_l)$. This expression is similar to the expression given by Ahn {\it et} al.\cite{ahn_PRX2020} using the Fermi golden rule. Notably, the injection conductivity is linearly proportional to the relaxation time and is finite only in the dirty limit. In addition, this is the Fermi sea response, and the corresponding real part of the response is proportional to the quantum metric. On the other hand, taking limits such as $1/\tau_\gamma \rightarrow 0$ and then $\omega_{\delta} \rightarrow 0$ the first term of Eq.~\eqref{eqn:GFs} becomes zero and the nonlinear conductivity becomes
\begin{align} \nonumber
\label{eqn:Wata}
&\sigma_{abc}^{do} (\omega_{\delta}; \omega_l,\omega_j) =\\[1ex]   &-i \lim_{\omega_{\delta} \rightarrow 0} \frac{2\pi e^3 }{\omega_{\delta}\hbar^2} \sum_{b,c} \sum_{m,p} \Delta_{mp}^{a} \mathcal{Q}_{mp}^{bc}
F_{mn}  \delta (\omega_{mp} - \omega_l).
\end{align}
This expression diverges at $\omega_\delta \rightarrow 0$ and is consistent with the calculation quoted recently using the kinetic approach~\cite{watanabe_PRX2021}. However, such divergence can be avoided by inserting the relaxation time factor. Further, here the real part of the response depends on the Berry curvature due to the imaginary factor in the expression. 

Note that Eqs.~\eqref{eqn:Naga} and \eqref{eqn:Wata} are obtained by taking limiting cases and yield different results for the injection current. One needs to be cautious before taking the limits as it may lead to different results. To avoid such confusion, we thoroughly provide the complete expression (Eq.~\ref{sigma_do}) of the injection conductivity without taking any limits which captures both the cases shown in the literature~\cite{ahn_PRX2020, watanabe_PRX2021}. 
Further, we observe that the resonance features survive only for the doped system, and the resulting peak is controlled by the relaxation time scale. This is ultimately a Fermi sea effect determined by the joint density of states and the band velocity difference. 

\subsubsection{Interband-Intraband Current}
We consider the current due to the off-diagonal contribution of the second-order density matrix depending on the diagonal part of the first-order density matrix. It reads
\begin{equation}
    j_a^{(2),od}(t) = -e \sum_{m}\sum_{\bm k} v_{pm}^{a} \rho_{mp}^{(2),od}(t).
\end{equation}
Substituting the expression for the interband-intraband density matrix component $\rho_{mp}^{(2),od}$ from Eq.~\eqref{eqn:do}, the interband-intraband nonlinear response becomes
\begin{equation}
\sigma^{od}_{abc}(\omega_{\delta}; \omega_j, \omega_l) = 
-  \frac{e^3 }{\hbar^2} \sum_{\mathcal{P}}\sum_{m,p} \omega_{mp}\mathcal{Q}_{mp}^{ab}  g_{0;\alpha}^{\omega_l} g_{mp;\gamma}^{\omega_{\delta}}  
\partial_c F_{mp}.
\label{sigma_do}
\end{equation}
The presence of the momentum derivative of the difference in the band occupation, $\partial_c F_{mp}$ dictates that Eq.~\eqref{sigma_do} is a Fermi surface effect. Although this current is a second-order effect, it does not simultaneously generate one- and two-photon absorption processes. Instead, it only gives either absorption process depending on the incident energy of two optical beams. To separate the resonant and non-resonant parts of the conductivity, it is helpful to employ the identity
\begin{equation}
\dfrac{\omega_{mp}}{1/\tau_\gamma -i (\omega -\omega_{mp})} = -i \left[ 1  - \dfrac{1/\tau_\gamma - i \omega}{1/\tau_\gamma -i (\omega -\omega_{mp})}\right].
\label{separation}
\end{equation}
Using the above relation we separate Eq.~\eqref{sigma_do} into two parts as $\sigma^{od}_{abc}(\omega_{\delta}; \omega_j, \omega_l) =\sigma^{od,I}_{abc}(\omega_{\delta}; \omega_j, \omega_l) +\sigma^{od,II}_{abc}(\omega_{\delta}; \omega_j, \omega_l) $. The first (non-resonant) part is to be
\begin{equation}
\sigma^{od,I}_{abc}(\omega_{\delta}; \omega_j, \omega_l) = \frac{e^3 }{2\hbar^2} \sum_{\mathcal{P}} \sum_{m,p} \Omega_{mp}^{ab} g_{0;\alpha}^{\omega_l} 
\partial_c F_{mp} .
\label{sigma_do_I}
\end{equation}
We emphasize that when writing Eq.~\eqref{sigma_do_I} we used the fact that $\Omega_{mp}$ is anti-symmetric and ${\mathcal G}_{mp}$ is symmetric in band index. Such simplification based on symmetry and anti-symmetry is only possible since there are no resonance factors with the band index ($g_{mp}$). This non-resonant part is known as the anomalous nonlinear response~\cite{rostami_PRB2018_nonlinear}. The other (resonant) part of the conductivity is given by
\begin{equation}
\sigma^{od,II}_{abc}(\omega_{\delta}; \omega_j, \omega_l) = - i \frac{e^3 }{\hbar^2} \sum_{\mathcal{P}}\sum_{m,p} \mathcal{Q}_{mp}^{ab}\partial_c F_{mp} \frac{g_{0;\alpha}^{\omega_l}g_{mp;\gamma}^{\omega_{\delta}}}{g_{0;\gamma}^{\omega_{\delta}}}.
\label{sigma_do_II}
\end{equation}
Eq.~\eqref{sigma_do_II} is one of the central results of this paper. This part of the current will show resonance behavior near the Fermi energy, which can be inferred from the derivative of the Fermi function. The optical field-dependent part is
\begin{equation} \label{optical_field}
\frac{g_{0;\alpha}^{\omega_l}g_{mp;\gamma}^{\omega_{\delta}}}{g_{0;\gamma}^{\omega_{\delta}}} = \dfrac{1}{1/\tau_\alpha -i \omega_l }\times  \dfrac{1/\tau_\gamma - i \omega_{\delta}}{1/\tau_\gamma -i (\omega_{\delta} -\omega_{mp}) }.
\end{equation}
In the clean limit, one obtains
\begin{equation} \label{clean}
g_{0;\alpha}^{\omega_l} g_{mp;\gamma}^{\omega_{\delta}} =i \dfrac{1}{\omega_l }\times  \dfrac{\omega_{\delta}}{\omega_{\delta} -\omega_{mp} }.
\end{equation}
and in such a case, this current diverges at the lower energy scale and approaches zero at higher energy. We emphasize that it is not right to consider a dc field limit of Eq.~\eqref{clean} after applying the clean limit. Instead, it is preferable to consider the dc limit from Eq.~\eqref{optical_field} and then consider the clean or dirty limit.
Notably, the resonant feature here will be observed at $\omega_\delta = \omega_{mp}$.

\subsubsection{Interband-Interband Current}
Finally, the substitution of the interband part of the first-order density matrix into the interband part of second-order $\rho_{mp}^{(2)}$ yields the interband-interband current. It is defined as
\begin{equation}
j_a^{(2),oo}(t) = -e \sum_{m,p} v_{pm}^{a} \rho_{mp}^{(2),oo}(t).
\end{equation}
We may refer to this current as purely an interband coherence current as it arises from interband geometric quantities, off-diagonal components of the velocity, and the joint density of states. Now, with [Eq.~\eqref{eqn:oo}] for $\rho_{mp}^{(2),oo}$ and using the definition $\mathcal{D}_{mp}^b = \partial_b  -i  \left( \mathcal{R}^{b}_{mm}  - \mathcal{R}^{b}_{pp} \right)$, the corresponding nonlinear conductivity part comes out to be
\begin{equation} \label{sigma_oo}
\ba
&\dps \sigma_{abc}^{oo}(\omega_{\delta}; \omega_j, \omega_l)  \\[3ex]
&\dps= -\frac{e^3}{\hbar^2} \sum_{\mathcal{P}}\sum_{m,p} \mathcal{R}^a_{pm} \omega_{mp} g_{mp;\gamma}^{\omega_{\delta}}  \Big[ {\mathcal D}_{mp}^b {\mathcal R}_{mp}^c F_{mp}g_{mp;\gamma}^{\omega_l} 
\\[3ex]
&\dps -i \sum_{n \neq (m,p)} \left( \mathcal{R}^{b}_{mn}{\mathcal R}_{np}^c F_{np}g_{np;\gamma}^{\omega_l}  -{\mathcal R}_{mn}^c \mathcal{R}^{b}_{np} F_{mn}g_{mn;\gamma}^{\omega_l}  \right) \Big].
\ea
\end{equation}
Further with the help of Eq.~\eqref{separation} the above mentioned conductivity can be written as $\sigma_{abc}^{oo}(\omega_{\delta}, \omega_j, \omega_l)=\sigma_{abc}^{oo,I}(\omega_{\delta}, \omega_j, \omega_l)+\sigma_{abc}^{oo,II}(\omega_{\delta}, \omega_j, \omega_l)+\sigma_{abc}^{oo,III}(\omega_{\delta}, \omega_j, \omega_l)$ by applying the covariant derivative separately to distinct factors. The last term of the above equation (Eq.~\ref{sigma_oo}) contributes only to the multi-band (more than two bands) systems. Here, the first part is given by
\begin{equation} \label{sigma_ooI}
\ba
&\dps \sigma_{abc}^{oo,I}(\omega_{\delta}; \omega_j, \omega_l)\\ &\dps= -\frac{e^3}{\hbar^2} \sum_{\mathcal{P}}\sum_{m,p} \mathcal{R}^a_{pm} \omega_{mp} g_{mp;\gamma}^{\omega_{\delta}} g_{mp;\gamma}^{\omega_l}F_{mp} {\mathcal D}_{mp}^b {\mathcal R}_{mp}^c .
\ea
\end{equation}
More simplifications are made using the sum rule~\cite{cook_NC2017} for the covariant derivative on the Berry connection. Accordingly,
\begin{equation}
\ba
{\cal D}^b_{mp}  {\mathcal R}^c_{mp}   =&\dps  - \dfrac{1}{i \omega_{mp}} \left[ \dfrac{v_{mp}^b \varDelta_{mp}^c + v_{mp}^c \varDelta_{mp}^b }{\omega_{mp}} \right.
-w_{mp}^{bc} \\[3ex]
&\dps \left.+ \sum_{n \neq (p,m)} \left(\dfrac{v_{mn}^c v_{np}^b}{\omega_{np}} - \dfrac{v_{mn}^b v_{np}^c}{\omega_{mn}} \right)\right].
\ea
\label{sum_rule}
\end{equation}
Here, we have defined $w_{mp}^{bc} = \langle m | \partial_b \partial_c {\cal H} | p \rangle$. From Eq.~\eqref{sum_rule}, it is straightforward that the last term does not contribute to the two-band model. 
More precisely, the corresponding conductivity for two band model can be written in the form as
\begin{equation}
\ba
\sigma^{oo,I}_{abc}  &\dps=  - \frac{e^3 }{\hbar^2 } \sum_{\mathcal{P}}\sum_{m,p}  \omega_{mp}  \mathcal{C}_{mp}^{abc}  g_{mp}^{\omega_{\delta}} g_{mp}^{\omega_l} F_{mp},
\ea
\label{j_shift_n}
\end{equation}
where $\mathcal{C}_{mp}^{abc} = \mathcal{R}_{pm}^{a}  \mathcal{D}^{b}_{mp}\mathcal{R}^c_{mp} = \Gamma_{mp}^{abc} + i \tilde{\Gamma}_{mp}^{abc}$ is the quantum geometric connection which is the sum of the quantum geometric quantities namely metric connection $\Gamma_{mp}^{abc}$ and symplectic connection $\tilde{\Gamma}_{mp}^{abc}$.
Originally, the quantum geometric connection $\mathcal{C}_{mp}^{abc}$ in the tangent subspace spanned by basis vectors $\hat{e}_{mn}^b$ stems from the inner product of the tangent basis vector $\hat{e}_{mp}^b$ and the derivative of such vectors $\nabla_b \hat{e}_{mp}^c$. Further, $\tilde{\Gamma}_{mp}^{abc}$ is directly related to the shift vector. To elaborate, consider the case of $a=c$ and write Berry connection as $\mathcal{R}_{mp}^a = |\mathcal{R}_{mp}^a|e^{i\phi_{mp}}$ with $\phi_{mp}$ a phase factor. With this, we get
\begin{align}
    \tilde{\Gamma}_{mp}^{abc} &= \vert {\mathcal R}^a_{mp} \vert^2  \partial_b  \phi_{mp}  - \left( \mathcal{R}^{b}_{mm} -  \mathcal{R}^{b}_{pp} \right)\vert {\mathcal R}^a_{mp}\vert^2~.
\end{align}
This expression is consistent with the shift vector definition~\cite{morimoto_SA2016}. The difference between Berry connections indicates the difference of shifted Bloch wave functions between conduction and valence band, and the momentum derivative of the phase factor $\partial_b  \phi_{mp}$ maintains the gauge invariance. Further, using Eq.~\eqref{separation} and expressing $g_{mp}^{\omega_l}$ with Sokhotski–Plemelj relation, it is straightforward to express the interband-interband response in the form of the shift response as shown in the literature~\cite{ahn_PRX2020,kim_PRB2017,watanabe_PRX2021}. 
The second part of $\sigma_{abc}^{oo}$ is
\begin{equation} \label{sigma_oo}
\sigma_{abc}^{oo,II}(\omega_{\delta}; \omega_j, \omega_l) 
= - \frac{e^3 }{\hbar^2} \sum_{\mathcal{P}} \sum_{m,p}  \omega_{mp}g_{mp}^{\omega_{\delta}}   g_{mp}^{\omega_l}  {\mathcal Q}^{ac}_{mp} \partial_b F_{mp}.
\end{equation}
This is the Fermi surface effect and arises due to the asymmetric Fermi surface in the momentum space and the band geometric quantities. To illustrate more, let us consider the case of low temperature. In this limit, the momentum derivative of the Fermi function approaches to $-\delta(\omega_{mp} - \mu) \partial \varepsilon/\partial {\bm k}$ having $\mu$ the chemical potential. If we perform the partial separation of $g_{mp}^{\omega_{\delta}} g_{mp}^{\omega_l}$ and then solve the $k$ integral, the response gives two resonant peaks at energy $\omega_{\delta} = 2\mu$ and $\omega_l = 2\mu$. Note that the corresponding response persists irrespective of geometry. Due to this feature, the associated current is known as the double resonant current. However, in the case of $\omega_{\delta} \rightarrow 0$ (the optical rectification process), one of the peaks disappears, and the induced effect is known as the resonant photovoltaic effect~\cite{bhalla_PRL2020}. Here, the integrand becomes proportional to $2\tau \partial \varepsilon/\partial k_a \delta(\omega_{mp} - \varepsilon_F)$ at energy scale $\omega_l = 2\mu$. Clearly, the effect originates due to the Fermi surface displacement and is finite for the doped systems.

The third part of the interband-interband response, due to the momentum derivative of the joint density of states, is
\begin{align}\nonumber \label{sigma_oo}
&\sigma_{abc}^{oo,III}(\omega_{\delta}; \omega_j, \omega_l)  
= \\[1ex]
&-\frac{e^3}{\hbar^2} \sum_{\mathcal{P}}\sum_{m,p} \omega_{mp} \mathcal{Q}^{ac}_{pm}  g_{mp;\gamma}^{\omega_{\delta}} [\partial_b g_{mp;\gamma}^{\omega_l}] F_{mp}.
\end{align}
The associated current is known as the higher-order pole current. This is finite only for the finite scattering time and vanishes at $1/\tau \rightarrow 0$. To elaborate on the point, we consider the imaginary part of $g_{mp;\gamma}^{\omega_{\delta}} [\partial_b g_{mp;\gamma}^{\omega_l}]$ which gives
\begin{equation}
\ba
&\dps {\rm Im}[g_{mp;\gamma}^{\omega_{\delta}} \partial_b g_{mp;\gamma}^\omega] 
= \\[2ex]
&\dps \frac{ (\partial_b \omega_{mp})  1/\tau_{\gamma}}{1/\tau_{\gamma}^2 + (\omega_{\delta}-\omega_{mp})^2}  \bigg[ \frac{2 (\omega_l -\omega_{mp})(\omega_{\delta} -\omega_{mp})}{[1/\tau_{\gamma}^2 + (\omega_l -\omega_{mp})^2]^2}\\[3ex]
&\dps + \left( \dfrac{2(\omega_l - \omega_{mp})^2}{[1/\tau_{\gamma}^2 + (\omega_{l} - \omega_{mp})^2]^2} - \dfrac{1 }{1/\tau_{\gamma}^2 +( \omega_l - \omega_{mp})^2 } \right)   \bigg].
\ea
\end{equation}
Clearly, the non-zero imaginary part of the resonance factor arises only through the incorporation of a finite scattering timescale due to disorder. 

\subsection{Scattering time scale}
Here we discuss the difference between the present and earlier treatment of the scattering time in the response. In earlier works, $1/\tau$ was typically added as an infinitesimally small imaginary part in the frequency $\omega \rightarrow \omega + i\eta$ having $\eta = 1/\tau$ to achieve convergence in the low-frequency response~\cite{passos_PRB2018_nonlinear}. Here, $\eta$ originates from causality via the slow switching of the perturbation and is generally set to zero to obtain results in the clean limit. However, this approach cannot be applied to systems where the interference of intraband and interband transitions play a pivotal role in generating the different components of the nonlinear response. In addition, the subtle difference between the two approaches becomes more important while going beyond the linear response regime to calculate the nonlinear response. 

In the approach of the addition of a small imaginary term, the resonance factor becomes  
\begin{equation}
    \frac{1}{\omega_l + \omega_j -  \varepsilon} \rightarrow \frac{1}{\omega_l + \omega_j - \varepsilon + i \eta}~.
\end{equation}
Nevertheless, one must be cautious while calculating the response where the scattering term contributes twice. The qualitative aspects can be captured correctly without taking the factor of $2$ in front of $\eta$ arising via the addition of an imaginary factor to two frequencies. 
Still, it may significantly affect the shape of resonances and yield different results around the resonances with two approaches. Such subtle issues can be avoided by considering the finite scattering term in the equation of motion as incorporated in the present study.

\section{Symmetry and Geometrical analysis}

\subsection{Symmetry Analysis}
\label{sec:symmetry}
In this section, we perform the symmetry analysis of the nonlinear optical currents induced by the optical field. We describe how the properties of the nonlinear currents are restricted by various symmetries such as the parity (or space inversion) ($\mathcal{P}$), time reversal ($\mathcal{T}$), inversion-time reversal ($\mathcal{P}\mathcal{T}$) symmetries. We begin by recalling basic symmetry arguments. Firstly under $\mathcal{P}$ unitary transformation, the position vector ${\bm r}$ changes sign to $-{\bm r}$; thus, momentum changes sign. In this case, the Bloch Hamiltonian follows the eigenvalue equation as
\begin{equation}
    \mathcal{P} \mathcal{H}({\bm k}) | u_{{\bm k}}^m \rangle =  \mathcal{H}(-{\bm k}) \mathcal{P}| u_{{\bm k}}^m \rangle,
\end{equation}
which gives the energy eigenvalues of the Bloch Hamiltonian that remain invariant on changing ${\bm k}$ to $-{\bm k}$:
\begin{equation}
\varepsilon_{m}({\bm k}) = \varepsilon_{m}(-{\bm k}).
\end{equation}
However, this symmetry is only preserved if the dispersion is an even function of the momentum. Further, the Bloch eigen function follows the relation
\begin{equation}
   \mathcal{P}| u_{m,{\bm k}} \rangle = |u_{-{\bm k}}^m\rangle.
\end{equation}
The Berry connection $\mathcal{R}({\bm k})$, for two band system which contains the momentum derivative of the eigen function satisfies the following relation under parity inversion
\begin{equation}
\mathcal{R}_{pm}^b ({\bm k}) = -\langle u_{-{\bm k}}^p | i {\bm \nabla} | u_{-{\bm k}}^m \rangle =  -\mathcal{R}_{pm}^b(-{\bm k}).
\end{equation}
However, the Berry curvature and the geometric tensor having the product of two Berry connection factors remain invariant under $\mathcal{P}$ symmetry, 
\begin{equation}
\ba
\Omega_{pm}^{bc}({\bm k}) &\dps = \Omega_{pm}^{bc}(-{\bm k}), \quad G_{pm}^{bc} ({\bm k}) = G_{pm}^{bc}(-{\bm k}).
\ea
\end{equation} 

Secondly, the time inversion symmetry ($\mathcal{T}$) is an anti-unitary transformation in which complex number changes to its conjugate. Here the momentum ${\bm k} \equiv i\partial_{\bm r}$ changes $-{\bm k}$ due to the sign flip of $i$. Under $\mathcal{T}$ symmetry, the energy eigenvalues remain invariant. However, the Bloch function follows
\begin{equation}
    |u_{-{\bm k}}^n\rangle^* = e^{i\phi(k)}|u_{{\bm k}}^n\rangle.
\end{equation}
Interestingly, the Berry connection, in addition to the sign change, also reverses the band index order as dictated below
\begin{equation}
\ba
\mathcal{R}_{pm}^b ({\bm k}) &\dps = -\mathcal{R}_{mp}^b(-{\bm k}).
\ea
\end{equation}
However, the band index criteria emerge only for multi-band systems. Further, due to the complex conjugate condition under time-reversal transformation, the geometric quantities satisfy
\begin{equation}
\ba
\Omega_{pm}^{bc}({\bm k}) &\dps= -\Omega_{pm}^{bc}(-{\bm k}), \quad G_{pm}^{bc} ({\bm k}) = G_{pm}^{bc}(-{\bm k}).
\ea
\end{equation}
It is to be noted that the Berry curvature, which is equal to $i[\mathcal{R}_{mp}^b,\mathcal{R}_{mp}^c]/2$ reverses sign and the quantum metric does not.

Thirdly, the combination of the parity and time-reversal symmetry ($\mathcal{P}\mathcal{T}$) properties yield
\begin{equation}
\ba
\mathcal{R}_{pm}^b ({\bm k}) &\dps = -\mathcal{R}_{mp}^b({\bm k}) \quad \Omega_{pm}^{bc}({\bm k}) = 0, \quad G_{pm}^{bc} ({\bm k}) = G_{pm}^{bc}({\bm k}).
\ea
\end{equation}

Using these symmetries, we can understand the different contributions of the current qualitatively. Under $\mathcal{P}$ symmetry, the electric field reverses the sign as $E \rightarrow -E$, and the current also changes sign $j \rightarrow -j$. In the nonlinear currents, the product of two electric fields preserves the sign, which ensures that the only component of the conductivity tensor should be zero, which changes the sign to fulfill the condition for current. 

First, the $\sigma_{abc}^{dd}$ contribution under $\mathcal{P}$ symmetry depends on the following quantities
\begin{equation}
v_{mm}^{0a} = - v_{mm}^{0a} ;~~ \partial_b \partial_c f_{m}^{(0)} = \partial_b \partial_c f_{m}^{(0)};~~  g_{0;\alpha}^{\omega_l}  = g_{0;\alpha}^{\omega_l},
\end{equation}
which flips the sign of the intraband-intraband response as $\sigma^{dd}_{abc} = -\sigma^{dd}_{abc}$. The other components of the response follow as
\begin{equation}
\sigma^{do}_{abc}= -\sigma^{do}_{abc} ; ~~ \sigma^{od}_{abc}= -\sigma^{od}_{abc} ; ~~\sigma^{oo}_{abc}= -\sigma^{oo}_{abc},
\end{equation}
due to the sign reversal of the velocity change $\Delta_{mn}^a$, shift of the Fermi function in the momentum space $\partial_c F_{mp}$, and Berry connection respectively.

Second under time-reversal $\mathcal{T}$ symmetry, the electric field does not change sign, but the current follows $j (= dP/dt) \rightarrow -j$. Thus, the conductivity tensor must be an odd function under time-reversal symmetry. Here,
\begin{equation}
    \mathcal{Q}_{mp} = \mathcal{Q}_{mp}^* ; ~~ g_{0;\alpha}^{\omega_l}g_{mp;\alpha}^{\omega_{\delta}} = [g_{0;\alpha}^{\omega_l}g_{mp;\alpha}^{\omega_{\delta}}]^*
\end{equation}
However, only contribution stemming from the Berry curvature (odd in nature) in $\sigma_{od}$, $\sigma_{do}$, and $\sigma_{oo}$ will be non-zero. 

Third under $\mathcal{P}\mathcal{T}$ symmetry, we have
\begin{equation}
    \mathcal{Q}_{mp} = \mathcal{Q}_{mp} ; ~~ g_{0;\alpha}^{\omega_l}g_{mp;\alpha}^{\omega_{\delta}} = [g_{0;\alpha}^{\omega_l}g_{mp;\alpha}^{\omega_{\delta}}]^*
\end{equation}
Here, the Berry curvature vanishes, and the quantum metric remains non-zero. Thus, the only contribution from the quantum metric $G_{mp}$ generates a finite nonlinear current.

\subsection{Spatial Geometrical analysis}
We analyze the different components of the conductivity tensor due to the application of the optical field in distinct directions. \\
{\bf Scheme - I}: When the optical field is applied in the $\hat{x}$-direction such as ${\bm E}(t) = \sum_{j}E_x^{\omega_j} e^{-i\omega_j t}$, the longitudinal current $j_x^{(2)}$ for the time-reversal symmetric system vanishes due to the vanishing Berry curvature $\Omega_{bc} = \Omega_{xx}$. However, it remains finite for the parity-time reversal symmetric system due to non-zero quantum metric $G_{xx}$. Conversely, the Berry curvature contributes to the transverse current $j_y^{(2)}$ due to the non-zero response components $\sigma_{od}$ and $\sigma_{oo}$. Note that the injection current due to $\Omega$ here is zero. \\
{\bf Scheme - II}: For the optical field as the superposition of the beams in $\hat{x}$ and $\hat{y}$-directions, ${\bm E}(t) = E_x^{\omega_j} e^{-i\omega_j t} + E_y^{\omega_l} e^{-i\omega_l t}$, the current along $\hat{x}$-direction is mainly contributed by the two tensor components $\sigma_{xxy}$ and $\sigma_{xyx}$. In the spatial geometry $xxy$, the resonant part of the interband-intraband conductivity or the anomalous conductivity vanishes, and the non-resonant part gives a finite value via the finite quantum metric. In addition, the other parts also contribute to the current through $G$. For $xyx$ geometry, the double resonant part $\sigma_{xyx}^{oo,II}$ and the higher order pole part $\sigma_{xyx}^{oo,III}$ are zero for the time reversal symmetric system due to vanishing antisymmetric Berry curvature $\Omega_{xx}$. Similarly for the nonlinear current $j_y^{(2)}$, we have $\sigma_{yxy}$, and $\sigma_{yyx}$ components.

\section{Applications}
\label{sec:applications}
In this section, we discuss the application of the general
kinetic approach for the optical currents developed in the present study to known models. Our focus in this
section is to establish the connection of the geometric quantities with the nonlinear response components using a few examples. However, the framework is formulated in this paper in a more general way that is appropriate for all systems. 
\begin{figure}[t]
    \centering
    \includegraphics[width=8.5cm]{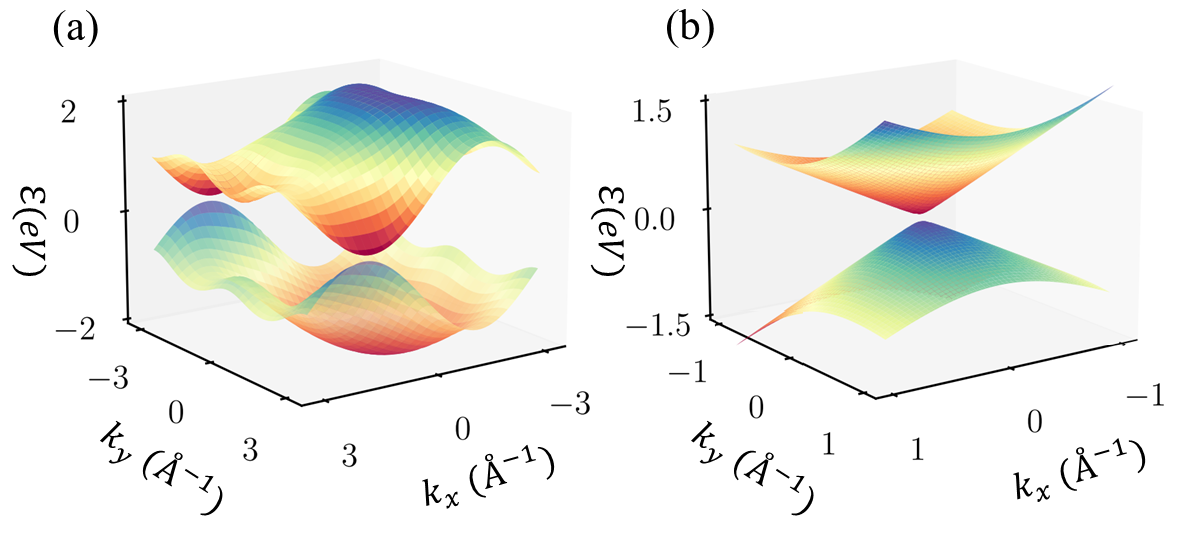}
    \caption{Schematic picture of the energy dispersion in the two-dimensional momentum plane for (a) Topological antiferromagnet CuMnAs and (b) Gapped tilted Weyl semimetal. Here, the momentum coordinates are labeled as $k_x$ and $k_y$, and the third dimension corresponds to the energy. We consider the parameter values in the units of eV as $\tilde t$=1.0, $t$ = 0.08, $\alpha_R$=0.8, $\alpha_D$=0.0, $h_{AFx}$=0.85, $h_{AFy}$=0, $h_{AFz}$=0 for CuMnAs system. For tilted Weyl semimetal, $t_x = 0.5$ eV\AA$^{-1}$, $\Delta = 0.05$ eV and $v$=1.0 eV\AA$^{-1}$.}
    \label{fig:dispersion}
\end{figure}

\subsection{Topological antiferromagnetic CuMnAs}
Here we consider the case of Dirac semimetal, which has attracted attention as a host of massless Dirac quasiparticles with two doubly degenerate bands in the momentum space~\cite{liu_Sc2014,liu_NM2014,yang_NC2014}. These doubly degenerate bands having band crossing between them generate four-fold degenerate Dirac points. However, such Dirac points are unstable and require symmetry protection~\cite{yang_NC2014}. Specifically, upon breaking either symmetry, such as time-reversal and parity, the double degeneracy of the bands is lifted, and the massless Dirac quasiparticles break down. This raises a natural question about the existence of such fermions in the absence of individual symmetry and in the presence of the combination of both the time-reversal and parity symmetry. Such a picture has been addressed by considering the example of a 2D material, CuMnY where Y is As or P, having the spin-orbit coupling that protects the band crossings in the Dirac semimetal~\cite{tang_NP2016}. In the paramagnetic phase, this material preserves the time-reversal
and parity symmetries. This results in the formation of Kramers pairs by each band. On the other hand, in the antiferromagnetic phase material breaks both $\mathcal{T}$ and $\mathcal{P}$ symmetry. However, it preserves the degeneracy due to the $\mathcal{P} \mathcal{T}$ symmetry which relates to the spin degrees of freedom. This exciting feature makes the antiferromagnetic phase of the CnMnAs an excellent choice to study transport effects such as the spin-orbit torque, spin Hall effect, and anomalous Hall effect~\cite{wadley_Sc2016, smejkal_PRL2017, maca_JMMM2012, smejkal_NRM2022, wang_PRL2021, liu_PRL2021}.  

\begin{figure}[t]
    \centering
    \includegraphics[width=9cm]{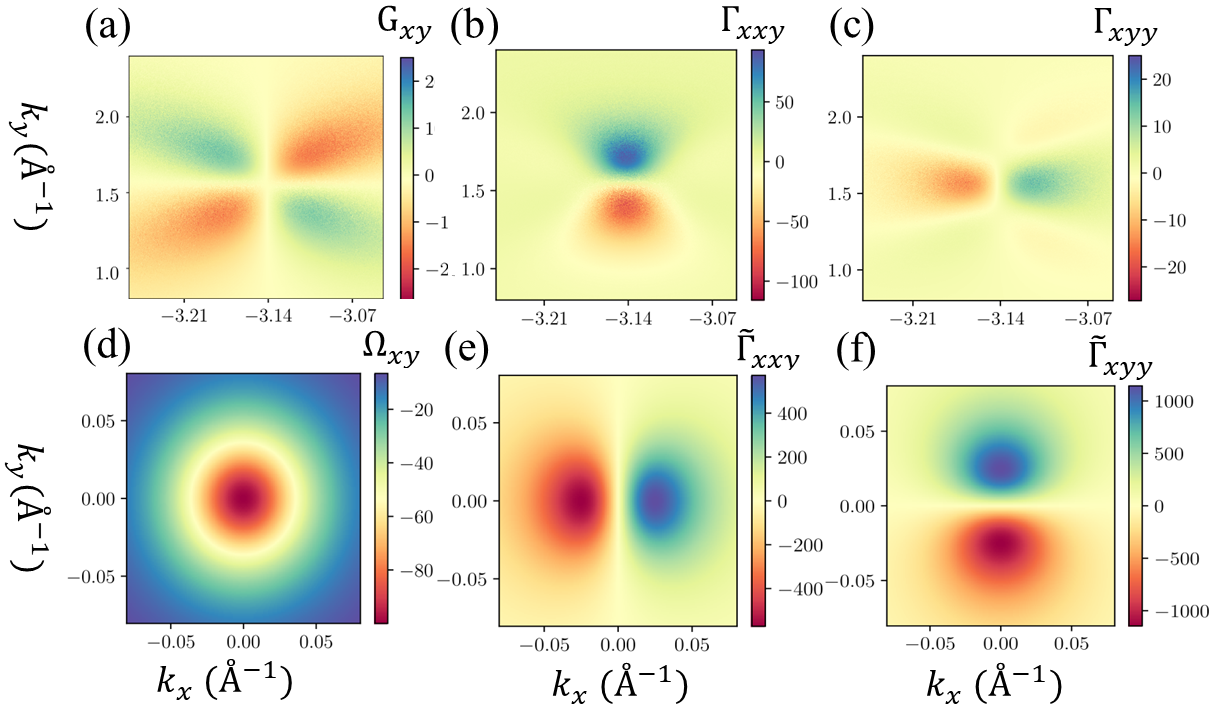}
    \caption{Distribution of the geometric quantities in the momentum space. Top panel: for Topological antiferromagnetic CuMnAs where (a) corresponds to the quantum metric, (b) and (c) to metric connection. Bottom panel: for thin film tilted Weyl semimetal where (d) refers to the Berry curvature, (e) and (f) to the symplectic connection.}
    \label{fig:GQ}
\end{figure}

The $\mathcal{P}\mathcal{T}$ symmetric topological antiferromagnetic CuMnAs material is described by the low energy 2D model Hamiltonian in the momentum space as \cite{Jungwirth_PRL_2017}
\begin{equation}
\ba \label{CuMnAs}
\mathcal{H}({\bm k}) &\dps = 
\begin{pmatrix}
\varepsilon_0 ({\bm k}) + {\bm h}_{\rm A}({\bm k}) \cdot {\bm \sigma} & V_{\rm AB}({\bm k})\\
V_{\rm AB}({\bm k}) & \varepsilon_0 ({\bm k}) + {\bm h}_{\rm B}({\bm k}) \cdot {\bm \sigma}
\end{pmatrix},
\ea
\end{equation}
where $V_{\rm AB}({\bm k}) = -2 \tilde t \cos (k_x/2) \cos (k_y/2)$ is the inter-sublattice hopping term having $\tilde t$ the first nearest neighbour hopping parameter, $\varepsilon_0({\bm k}) = -t \left[\cos (k_x) + \cos(k_y)\right]$ is the intra-sublattice hopping term with parameter $t$ as the second-nearest neighbour and $\sigma_i$ represent the Pauli matrices for spin. 
Further, the quantity ${\bm h}_{\rm A}({\bm k})$ for the sublattice $A$ that includes the antiferromagnetic (AF) magnetization field and the spin-orbit coupling (SOC) term ${\bm h}_{\rm A}({\bm k}) = {\bm h}_{\rm AF} + {\bm h}_{\rm SOC} ({\bm k})$ is defined like
\begin{equation}
\ba
{\bm h}_{\rm A}({\bm k}) =
\begin{pmatrix}
h_{\rm AF}^x - \alpha_{\rm R} \sin (k_y) + \alpha_{\rm D} \sin (k_y) \\
h_{\rm AF}^y + \alpha_{\rm R} \sin (k_x) + \alpha_{\rm D} \sin (k_x) \\
h_{\rm AF}^z
\end{pmatrix},
\ea
\end{equation}
having $\alpha_R$ and $\alpha_D$ as the spin-orbit coupling coefficients and for sublattice $B$, ${\bm h}_{\rm B}({\bm k})=-{\bm h}_{\rm A}({\bm k})$. The energy eigenvalues corresponding to the Hamiltonian Eq.~\eqref{CuMnAs} are 
\begin{equation}
\varepsilon({\bm k}) = \varepsilon_0 \pm \sqrt{V_{\rm AB}^2 + h_{{\rm A}x}^2 + h_{{\rm A}y}^2 + h_{{\rm A}z}^2}~.    
\end{equation}
Here, (+) sign is for the conduction band and (-) for the valence band. Further, the dispersion $\varepsilon({\bm k}) \neq \varepsilon(-{\bm k})$ due to the broken particle-hole symmetry by $\varepsilon_0$. The schematic picture of the dispersion is shown in Fig.~\ref{fig:dispersion} (a), and the corresponding geometric quantities are shown in Fig.~\ref{fig:GQ} (a)-(c). Here, the band crossing at the Dirac points is protected due to the glide planer symmetry~\cite{smejkal_PRL2017}. Further, the quantum geometric quantities Berry curvature and symplectic connection are zero in this system due to symmetry arguments. However, the other quantities quantum metric $G_{ab}$ and symplectic connection $\Gamma_{abc}$ are non-zero where $(a,b,c) \in (x,y)$. To demonstrate the nature of these quantities, we have shown a few components in the top panel of Fig.~\ref{fig:GQ} where we have considered the antiferromagnetic magnetization field along $\hat{x}$-direction and zero in other directions.

\begin{figure}[t]
    \centering
    \includegraphics[width=8.5cm]{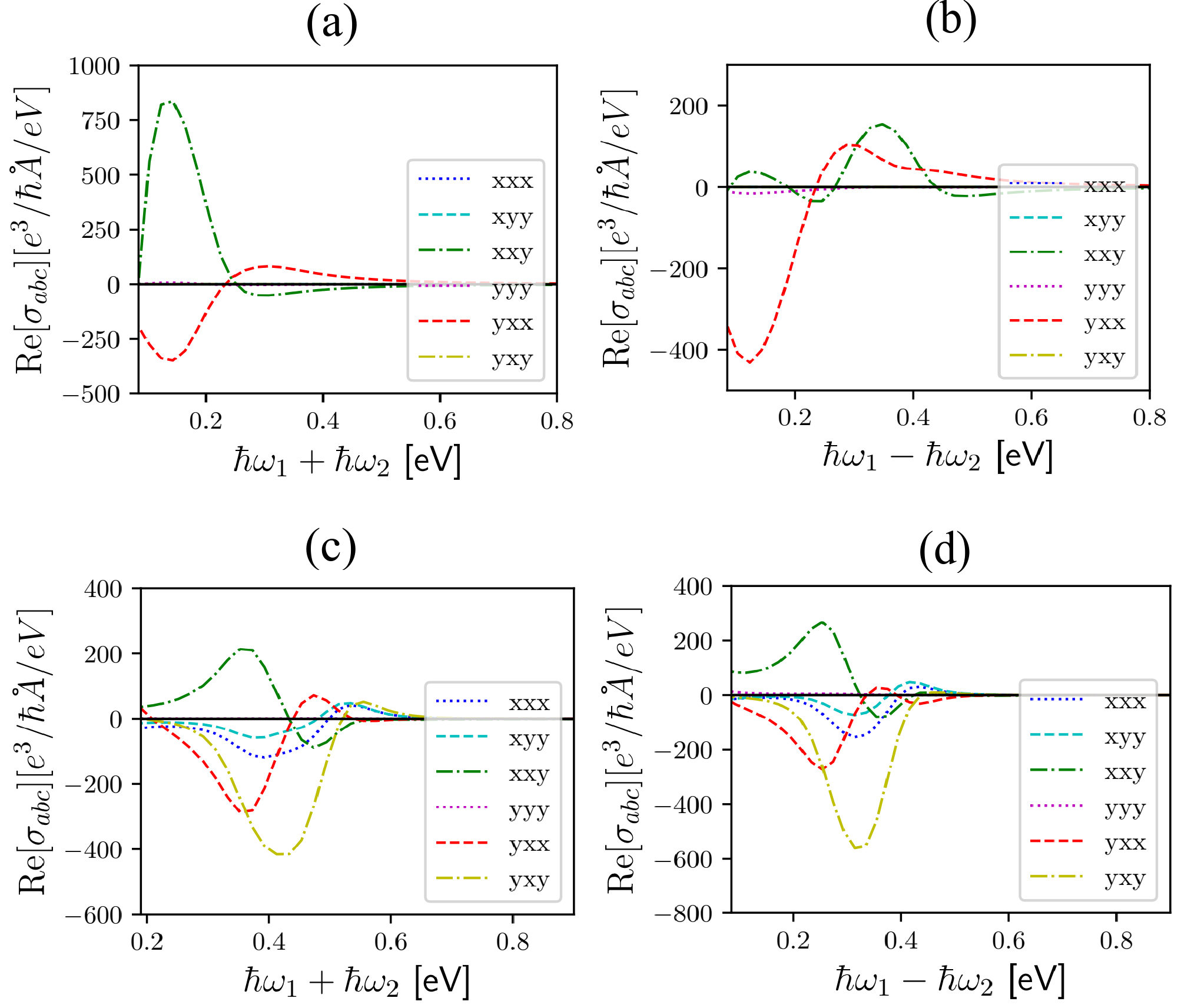}
    \caption{Different components of the second-order response due to the two beams having frequency $\omega_1$ and $\omega_2$. Column 1: (a) and (c) depict the sum frequency generation phenomenon, and Column 2: (b) and (d) show the difference in frequency generation effect. The output is obtained at fixed frequency $\omega_2 = 0.1$ eV, and the chemical potential $\mu = 0.2$ eV, $\tau=1$ ps and temperature $T=10$ K, but varies $\omega_1$. Figures (a) and (b) correspond to the CuMnAs where we set the hopping $t=0.08$ eV and $\tilde t=1$ eV. The other parameters are $\alpha_{\rm R}=0.8, \alpha_{\rm D}=0$ and ${\bm h}_{\rm AF}=(0.85,0,0)$ eV. Figures (c) and (d) refer to the thin film tilted Weyl semimetal where we consider the gap $\Delta = 0.05$ eV and the tilt $t_x = 0.1$ eV\AA and temperature $T = 1$K.}
    \label{fig:Sigma}
\end{figure}

In the parity time-reversal symmetric CuMnAs system, eight components of the nonlinear conductivities contribute to the second-order current, in general. However with the finite magnetization field along $\hat{x}$-direction, the components with odd number of spatial $x$ indices such as $\sigma_{xxx}$, $\sigma_{xyy}$, $\sigma_{yyx}$ and $\sigma_{yxy}$ vanish. Thus, we left with $\sigma_{yxx}$, $\sigma_{xxy} = \sigma_{xyx}$ and $\sigma_{yyy}$. The behavior of these components with the incident beam frequency at the low temperature is shown in Fig.~\ref{fig:Sigma}(a) and (b). Here we fix the frequency of one incident beam $\omega_2$ and tune the frequency of another beam $\omega_1$, while the chemical potential is kept at $\mu = 0.2$ eV and the scattering time scale $\tau = 1$ ps. We observe that the total response, a sum of different components such as dd, do, od, and oo is mainly contributed by the geometric quantities $G_{ab}$ and $\Gamma_{abc}$. In CuMnAs, we find the following features. (i) The absorption peaks are generated at energies around $\mu$ and $2\mu$. The observed behavior represents interference between the Fermi surface (i.e., the momentum derivative of the Fermi distribution function) and the Fermi sea effects. Note that the deviation in peaks is due to the absence of particle-hole symmetry in the considered system. (ii) The nonlinear conductivity $\sigma_{xxy}$ is opposite in sign to the other components. This arises due to the opposite sign of the $\sigma^{do}$, $\sigma^{od}$ and $\sigma^{oo,I}$ stemming from the nature of the factors $G_{bc}v_a$, $G_{ab}v_c$ and $\mathcal{C}_{abc}$ respectively.

\subsection{Thin film tilted Weyl semimetal}
Secondly, we consider the tilted Weyl semimetal, which is a three-dimensional topological semimetal in general~\cite{potter_NC2014, burkov_NM2016}. Here the conduction and valence bands touch each other at the Weyl nodes having opposite chirality. Further, this material shows the phase transition from topological to trivial by tuning the gap controlled by an out-of-plane component of the momentum~\cite{culcer_PRB2017}. In the case of an ultrathin film of the Weyl semimetal, the out-of-plane component of the momentum is quantized, then the system becomes a two-dimensional fermion system~\cite{lu_PRB2010, abanin_PRL2011}. Further, this quantized component results in the mass or the gap between the bands, leading to intriguing quantum transport effects such as anomalous Hall effect, planar Hall effect, and weak localization and anti-localization effects~\cite{burkov_PRL2014, trescher_PRB2015, ferreiros_PRB2017, liu_PRB2017, ma_PRB2019}.

The effective Hamiltonian for the time-reversal symmetry broken tilted Weyl semimetal around a Weyl point~\cite{ma_PRB2019} is
\begin{equation}
    \mathcal{H}({\bm k}) = v {\bm k} \cdot {\bm \sigma} + {\bm t} \cdot {\bm k} \sigma_0 + \Delta \sigma_z. 
\end{equation}
where the first term represents the spin-orbit coupling term having $v$ as the effective velocity in units of eV m s$^{-1}$, $t$ is the tilt vector and $\Delta$ refers to the gap which distinguish thin film Weyl semimetal from the topological insulator. The energy eigenvalues are
\begin{equation}
    \varepsilon({\bm k}) = t_x k_x + t_y k_y \pm \sqrt{v^2 k^2 + \Delta^2}.
\end{equation}
Here, the tilt term breaks the time-reversal and inversion symmetry due to the linear momentum factor. Such tilt term does not affect the eigenvectors, hence the topology of the system. However, it affects the response of the system. In addition, the type of Weyl semimetal is defined by $|t| < v$ (Type-I) and $|t| > v$ (Type-II). The corresponding dispersion for the Type-I Weyl semimetal is shown in Fig.~\ref{fig:dispersion} (b), and the geometric quantities are shown in Fig.~\ref{fig:GQ} (d)-(f). Here, the gap between conduction and valence band is $2\Delta$. Without tilt, the band dispersion becomes identical to the topological insulator, with the mass term smaller than the spin-orbit coupling term. In the opposite case, i.e., at the large mass, it behaves as a massive fermion system. In addition, all the quantum geometric quantities are non-zero, and a few are shown in Fig.~\ref{fig:GQ}. The Berry curvature is finite only due to the presence of the gap around the Dirac point and vanishes at $\Delta = 0$.

As distinct from the antiferromagnetic CuMnAs system, all eight nonlinear response components contribute to the dynamical current for the thin film tilted Weyl semimetal. However, these reduce to six due to symmetrical properties for the tensor components such as $\sigma_{xyx} = \sigma_{xxy}$, and $\sigma_{yxy} = \sigma_{yyx}$. These are mainly dictated by the quantities $G_{ab}$, $\Omega_{ab}$, $\Gamma_{abc}$, and $\tilde{\Gamma}_{abc}$. The behavior of the nonlinear response components is shown in Fig.~\ref{fig:Sigma}(c) and (d). Here, we observe the following features. (i) The occurrence of the two absorption peaks, one at $\hbar\omega_\Sigma = 2\mu$ and the other at energy $\hbar\omega_\Sigma = 2\mu \pm \hbar\omega_2$. Note that the peak corresponding to the red and green curves is shifted as it is influenced by the contribution stemming from both finite Berry curvature and quantum metric. (ii) The generation of the resonant behavior of the response happens due to the finite Fermi surface effect. (iii) The $\sigma_{yxy}$ yields a larger magnitude than other nonlinear tensor components due to the stronger $\sigma^{oo,II}$ contribution stemming from the shifted Fermi surface in the momentum space along $\hat{x}$-direction because of the tilt $t_x$ and the quantum metric.

Experimentally, the presented results for the SFG and DFG are significant in terms of measurement geometry and doping. First, these nonlinear signals can be measured by invoking the measurement geometry of the nonlinear response in the distinct direction of the applied field, i.e., along and perpendicular to the field. In CuMnAs, one can have nonlinear current for the particular geometry, i.e., $j_{y}^{(2)} = \sigma_{yxx}E_0^2 \cos^2\gamma $, where $\gamma$ is a polarization angle made by applied field along $x$-axis. This results in the maximum current at $\gamma =0$. However, the current $j_{x}^{(2)} = \sigma_{xxy}/2 E_0^2\sin2\gamma$ yields more value at $\pi/4$ polarization angle. On the other hand, in thin film tilted Weyl semimetal, the nonlinear current can be obtained irrespective of the polarization angle as it contributes to all response components. Along $\hat{x}$-direction, the current follows $j_x^{(2)} = [\sigma_{xxx} \cos^2\gamma + \sigma_{xyy} \sin^2\gamma + \sigma_{xxy} \sin2\gamma] E_0^2$ and along $\hat{y}$-direction, $j_y^{(2)} = [\sigma_{yyy} \sin^2\gamma + \sigma_{yxx} \cos^2\gamma + \sigma_{yxy}\sin2\gamma] E_0^2$.  Second, the strength of the nonlinear current can be tuned with the chemical potential or by doping. By taking into account the Fermi level inside the band, the Fermi surface terms such as the resonant (a subpart of interband-intraband) and double resonant (a subpart of interband-interband) strengthen the peak value of the second-order response, hence the nonlinear current. 

\section{Summary}
\label{sec:summary}

We have systematically developed a general platform for evaluating the nonlinear response of a crystal to an oscillating electric field or laser field by taking into account the interband and intraband counterparts of the density matrix. In the linear regime, the intraband part of the response is captured by the band-diagonal component of the density matrix and the interband part by off-diagonal component which are responsible for the linear longitudinal conductivity, current-induced spin polarizations in spin-orbit coupled systems, anomalous Hall, and spin-Hall effects~\cite{nagaosa_RMP2010, inoue_PRB2004, culcer_PRL2007, culcer_2011}. However, on going beyond the linear regime by expanding the density matrix in terms of the external stimuli, it is not trivial to express the interband and intraband components directly. These are interconnected to each other and lead to distinct contributions to the nonlinear currents, such as intraband-intraband, intraband-interband, interband-intraband, and interband-interband. Further, the corresponding nonlinear current gives a significant contribution on account of the finite Fermi surface, which was not discussed earlier to the best of our knowledge. 

We employ our theory to describe the phenomena of sum frequency and difference frequency summation, which lead to the second-harmonic and rectification effect as a special case, respectively. Similarly, we identified the fundamental connection between the geometric quantum quantities and the nonlinear response. Based on the connection, we showed how the fundamental symmetries play a significant role in examining the physical origin of the different components of the nonlinear currents. 
An interesting and important fact is that the nonlinear optical currents are dominated due to the interband coherence contribution. We highlight the contribution of the individual part of the nonlinear optical current that provides insightful information. First, the intraband-interband current, known as the injection current, is calculated earlier in specific regimes of frequency and scattering time scale within the Fermi Golden rule, which lacks the detailed behavior of the particular current. Here, we calculated the explicit expression of the injection current without considering assumptions applicable to all regimes of interest. Second, we naively express the interband-intraband response in two parts. The non-resonant part corresponds to the well-known anomalous nonlinear current and is non-zero only if the Berry curvature is finite. On the other hand, the resonant part yields finite value in all systems and participates to have an absorption peak in the nonlinear response. Third, the interband-interband current leads to the shift, double resonant and higher-order pole sum frequency, and difference frequency summation nonlinear currents. We demonstrated the whole analysis for CuMnAs and thin film Weyl semimetal systems.  

Further, our theory considers the transport and optical responses on an equal footing, reflecting their interplay in the second-order response. The method developed in this work serves as a key tool to examine the intrinsic and extrinsic contributions of nonlinear currents obtained on general and fundamental grounds, and yield physical insight into the behavior of fermions. 

\acknowledgments

P.B. thanks SRM-AP for providing a high-performance computing facility. DC is supported by the Australian Research Council Centre of Excellence in Future Low-Energy Electronics Technologies, project number CE170100039. 

%

\end{document}